\def\sver{0}

\ifnum\sver=1
    \documentclass[9pt,twocolumn,twoside,lineno,table]{pnas-new-mod}
    \templatetype{pnasresearcharticle} 
\else
    \documentclass{article}
    \RequirePackage[numbers,sort&compress,merge,round]{natbib}
    \usepackage{hyperref,tabularx,colortbl,fullpage}
    \usepackage[table]{xcolor}
    \usepackage{amsfonts}
    \usepackage{setspace}
\fi

\usepackage{textcomp}  
\usepackage{ifdraft}
\usepackage{multirow}
\usepackage{graphicx}
\usepackage{booktabs}
\usepackage{fontawesome}

\newcommand{\mcrot}[4]{\multicolumn{#1}{#2}{\rlap{\rotatebox{#3}{#4}~}}}

\newcommand*{\twoelementtable}[3][l]%
{%
    \begin{tabular}[t]{@{}#1@{}}%
        #2\tabularnewline
        #3%
    \end{tabular}%
}

\ifdraft{\newcommand{\authnote}[3]{{\color{#3} {\bf  #1:} #2}}}{\newcommand{\authnote}[3]{}}

\newcommand{\onote}[1]{\authnote{Or}{#1}{blue}}

\newcommand{\chk}{\faCheck}
\newcommand{\x}{\faTimes}
\newcommand{\questionmark}{\faQuestion}
\newcommand{\minus}{\faMinus}
\newcommand{\volatile}{\faLineChart}

\title{Revisiting the Properties of Money\footnote{The opinions expressed in this article are the sole responsibility of the authors and should not be interpreted as reflecting the views of Sveriges Riksbank.}}

\ifnum \sver=1
    \author[a,1,2]{Isaiah Hull}
    \author[b,1]{Or Sattath} 
    
    \affil[a]{Research Division, Sveriges Riksbank, Stockholm, Sweden}
    \affil[b]{Department of Computer Science, Ben-Gurion University, Beersheba, Israel}
    
    \leadauthor{Hull and Sattath} 

    \significancestatement{The functions-and-properties of money framework was introduced in the late 1800s as a means of describing physical currencies, such as commodity money, metallic coins, and paper money. A form of physical money was considered to be good if it was cognizable, durable, divisible, portable, stable, fungible, and acceptable. This paper provides an update to this framework that incorporates the properties of digital forms of money, such as reversibility, backup, throughput, and local verifiability. It draws from both the economics and computer science literatures, and may provide useful input for both the analysis of currency competition and the design of digital money.}
    
    \equalauthors{\textsuperscript{1}I.H. contributed equally to this work with O.S.}
    \correspondingauthor{\textsuperscript{2}Address correspondence to e-mail: isaiah.hull@riksbank.se}
    
    
    \dates{This manuscript was compiled on \today}
    \doi{\url{www.pnas.org/cgi/doi/10.1073/pnas.XXXXXXXXXX}}
\else
    \usepackage{authblk}
    \author[1]{Isaiah Hull\footnote{Correspondence Address: Research Division, Sveriges Riksbank, SE-103 37, Stockholm, Sweden. Email: isaiah.hull@riksbank.se. Tel: +46 076 589 0661. Fax: +46 8 0821 05 31.}}
    \author[2]{Or Sattath}
    \affil[1]{Research Division, Sveriges Riksbank, Stockholm, Sweden}
    \affil[2]{Department of Computer Science, Ben-Gurion University, Beersheba, Israel}
\fi

\ifnum\sver=1
    \begin{abstract} 
    The properties of money commonly referenced in the economics literature were originally identified by Jevons~\cite{Jev76} and Menger~\cite{Men1892} in the late 1800s and were intended to describe physical currencies, such as commodity money, metallic coins, and paper bills. In the digital era, many non-physical currencies have either entered circulation or are under development, including demand deposits, cryptocurrencies, stablecoins, central bank digital currencies (CBDCs), in-game currencies, and quantum money. These forms of money have novel properties that have not been studied extensively within the economics literature, but may be important determinants of the monetary equilibrium that emerges in the forthcoming era of heightened currency competition. This paper makes the first exhaustive attempt to identify and define the properties of all physical and digital forms of money. It reviews both the economics and computer science literatures and categorizes properties within an expanded version of the original functions-and-properties framework of money that includes societal and regulatory objectives.
    \end{abstract}
    \pagebreak
\fi

\begin{document}
 
\maketitle
\ifnum \sver=1
    \thispagestyle{firststyle}
    \ifthenelse{\boolean{shortarticle}}{\ifthenelse{\boolean{singlecolumn}}{\abscontentformatted}{\abscontent}}{}
\else
    \begin{abstract} 
    \noindent The properties of money commonly referenced in the economics literature were originally identified by Jevons~\cite{Jev76} and Menger~\cite{Men1892} in the late 1800s and were intended to describe physical currencies, such as commodity money, metallic coins, and paper bills. In the digital era, many non-physical currencies have either entered circulation or are under development, including demand deposits, cryptocurrencies, stablecoins, central bank digital currencies (CBDCs), in-game currencies, and quantum money. These forms of money have novel properties that have not been studied extensively within the economics literature, but may be important determinants of the monetary equilibrium that emerges in the forthcoming era of heightened currency competition. This paper makes the first exhaustive attempt to identify and define the properties of all physical and digital forms of money. It reviews both the economics and computer science literatures and categorizes properties within an expanded version of the original functions-and-properties framework of money that includes societal and regulatory objectives. 
    \end{abstract}
    \pagebreak
\fi

\ifnum \sver=1
\dropcap{T}echnological
\else 
\section{Introduction}
Technological
\fi progress has historically enabled the development of new forms of money with novel and enhanced properties~\cite{Jev76, Men1892, Met12, Wil13, Har17}. The introduction of coins and paper money, for instance, improved portability and cognizability relative to commodity money. Private bank money offered the possibility to earn interest and (eventually) transact digitally. Cryptocurrencies, such as Bitcoin, provided censorship resistance. Central bank digital currencies, which are under research and development at an increasing number of central banks ~\cite{BHW20, BH19}, promise to restore public money, but in a digital form. And quantum money, which has been theoretically studied but is not yet technically feasible, could reproduce the properties of cash, but with improved unforgeability guarantees and the ability to transact digitally.\footnote{See Hull et al. \cite{HSDW20} for an overview of progress in the theoretical and experimental development of quantum money.}

\begin{figure}[h!]
\begin{center}
\ifnum \sver=0
\includegraphics[width=0.70\textwidth]{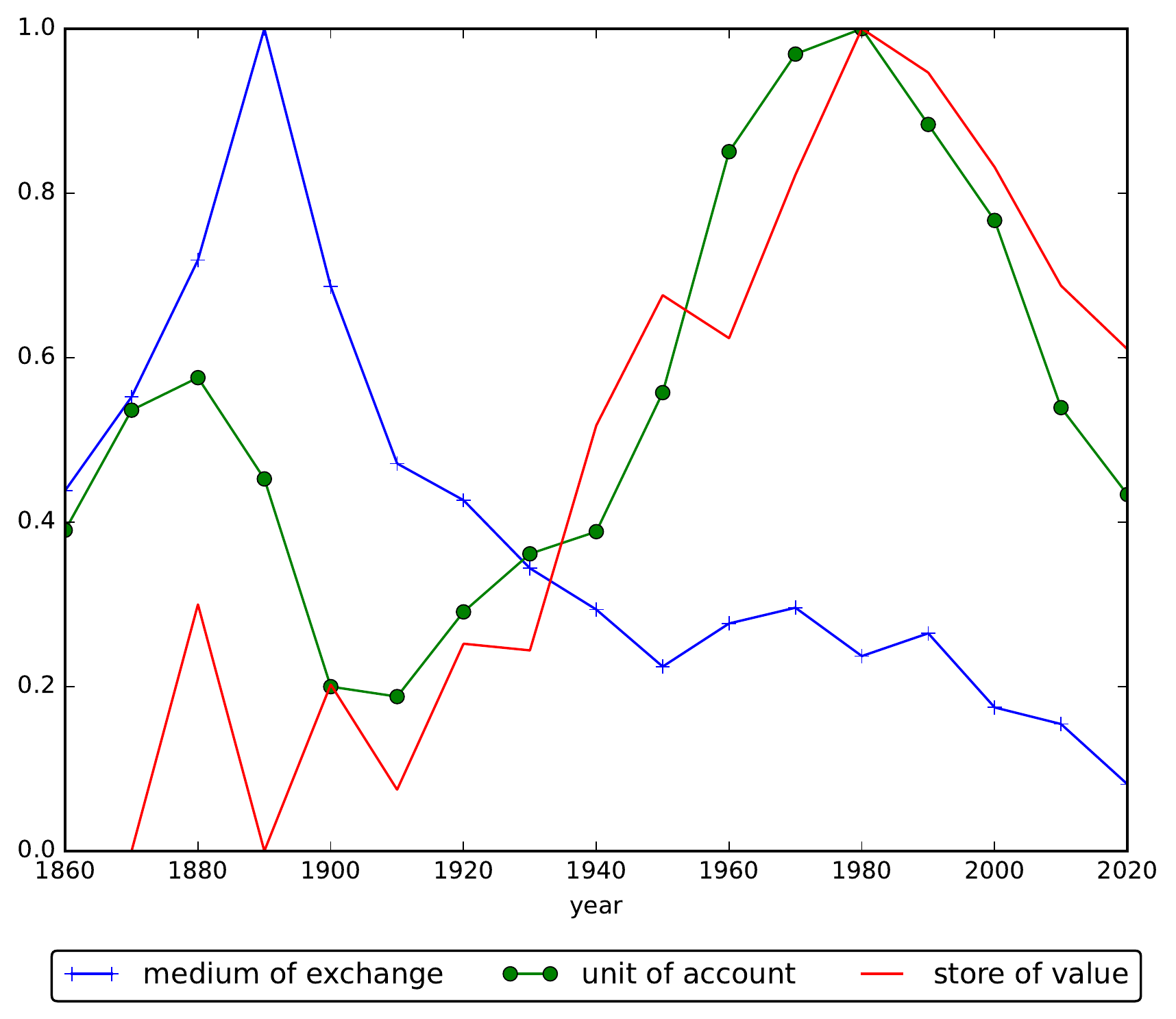}
\else
\includegraphics[width=0.49\textwidth]{Fig/money_ngrams.pdf}
\fi
\caption{The figure above shows phrase frequencies for the different functions of money in journal articles and books for the 1860-2020 period. Each series is normalized by the n-gram count for ``money'' and is then divided by its maximum value. The n-gram count data was generated by JSTOR Constellate.}
\label{fig:money_ngrams}
\end{center}
\end{figure}

While digital forms of money are now the preferred medium of exchange in many countries \cite{KH19}, the terminology used to describe money is still largely derived from foundational texts on physical currency, such as Jevons~\cite{Jev76} and Menger~\cite{Men1892}. Furthermore, the academic discussion of money's functions that followed these texts appears to have peaked prior to the development of digital currencies, as illustrated in Figure \ref{fig:money_ngrams} in Section \ref{sec:figures}. Consequently, many concepts that are routinely used in the modern literature on money were crystallized prior to the digital era.

Our intention is to update the standard framework for describing money by incorporating the properties of digital forms of money. To construct an exhaustive list of such properties, we not only review the economics literature, but also examine the parallel computer science literature, which approaches the properties of digital forms of money from a design perspective, focusing on what is achievable given a set of technical constraints. We also evaluate the performance of a selection of broad categories of money with respect to each of these properties in Table \ref{tab:properties_of_money}. This update to Jevons~\cite{Jev76} and Menger~\cite{Men1892}, which builds on recent work on the properties of money \cite{BG17,And18,BJL19,Eic19,BN19}, should have value for those doing research on CBDCs, cryptocurrencies, and digital payment schemes.

Part of the motivation for revisiting the properties of money is to provide better framing for the current period of rising currency competition, which follows an extended era of dominance by \hyperref[par:public]{public currencies} \cite{Kro11}. Whereas traditional forms of competition centered around physical proximity and macroeconomic integration \cite{BJL19}, emerging forms may center around less familiar concepts, such as \hyperref[par:throughput]{throughput}, \hyperref[par:latency]{latency}, and \hyperref[par:smart-contracts]{smart contracts}. Competition may happen within a set of uniform currencies, such as cash and bank deposits, or across non-uniform currencies, such as the U.S. dollar (USD) and Bitcoin. Our discussion of currency competition will adopt an inclusive definition that incorporates both.

The framework proposed in this paper could also be used to study the trade-offs inherent in money design choices, such as those discussed in Agur et al. \cite{AAD19} and Ferarri et al. \cite{FMS20}. Selecting one set of properties will necessarily entail excluding others. Consequently, placing too much emphasis on a property that is not broadly demanded (or is demanded by regulators, but not consumers) may result in a form of money that underperforms in a currency competition. One clear example of such a trade-off is the choice between \hyperref[par:untraceability]{untraceability} and anti-money laundering (AML) compliance. A less obvious trade-off is between 
\hyperref[par:localVerifiability]{local verifiability}, which is a form of forgery detection that does not require a third party, and the ability to secure against human and technical errors by performing \hyperref[par:backup]{backup}.

The continued relevance of public money in the 21st century may depend on how well central banks navigate these trade-offs \cite{BL17,BJL19,BSU19,CGM21}. In the previous round of currency competition, cash declined in use relative to private bank deposits \cite{Kro11, KH19}, suggesting that central banks were either incapable or uninterested in retaining control over the medium of exchange. In the emerging round of currency competition, the stakes may be even higher. Widespread adoption of a currency that is not uniform with a country's public money, such as a cryptocurrency or another central bank's CBDC (digital dollarization), could result in the loss of control over both the medium of exchange \textit{and} unit of account, as well as the inability to conduct monetary policy \cite{BJL19}. Many central banks appear to have concluded that it will be necessary to issue a form of money that is \textit{digital} in order to counter these threats \cite{BHW20, BH19}; however, no consensus exists on which other properties are necessary to remain competitive. Furthermore, it remains unclear whether a central bank would even want a CBDC to be truly competitive, as this might risk substantial disintermediation \cite{KK19}.

The return to an era of currency competition raises many regulatory and policy concerns; however, it also offers the possibility of improving money by incorporating the latest relevant technological advances, and extending the set of available regulatory and policy tools. Some have also argued that currency competition is needed to discipline central banks \cite{Hay76, MS05}. Others claim that the increase in competition from demand deposits has already resulted in institutional improvements \cite{Kro11}. The issuance of new forms of money could also lead to improvements in the measurement of monetary aggregates and an improved toolset for tracking consumption in real time during crises.\ifnum \sver=0 An expansion in the set of viable currencies could also have distributional implications by allowing otherwise marginalized and unbanked persons to transact digitally.
\fi

\paragraph{Organization.} %
First, we introduce the expanded set of properties of money and provide a definition for each. We sort them according to the functions introduced in Jevons~\cite{Jev76} and Menger~\cite{Men1892} and commonly cited in the literature: 1) medium of exchange, 2) standard of deferred payment, 3) store of value, and 4) unit of account. We also include a separate category for properties that enhance a societal or regulatory function. In the cases where a property affects multiple functions, we categorize according to the primary function. In addition to defining each property, we also examine the extent to which it is present in a set of broad categories of money in Table \ref{tab:properties_of_money}. 
\ifnum \sver=0 
Next, we discuss a selection of properties that apply to pairs or groups of currencies, rather than individual currencies.
\fi
Finally, we conclude with a discussion of the implications of CBDC and private currency design choices.


\newlength\nesw
\settowidth{\nesw}{$\nearrow$}
\def\neswarrow{\nearrow\hspace{-\nesw}\swarrow}

\newlength\nwse
\settowidth{\nwse}{$\nwarrow$}
\def\nwsearrow{\nwarrow\hspace{-\nwse}\searrow}

\ifnum\sver=1
\section*{Properties of Money}
\else 
\section{Properties of Money}
\fi

This section provides an update to the lists of monetary properties originally defined in Jevons~\cite{Jev76} and Menger~\cite{Men1892}, which are still frequently referenced, but were only intended to describe physical forms of money, such as commodity money or metallic coins. It draws from both the economics and computer science literatures, and provides an evaluation of several broad categories of money within this framework. For each property, we attempt to identify the function to which it corresponds. In cases where there are multiple functions, we categorize according to the primary function.

In general, we attempt to use a positive framing for each property. For example, we use \textit{low} pecuniary transaction cost as a property, rather than pecuniary transaction cost. In some cases, however, the properties we consider are not positive in an absolute sense, but may be desirable in the context of a specific design goal. Consider, for example, reversibility, which is the property that a transaction can be canceled under certain conditions. In some settings, the buyer's protection is the most important consideration and, thus, reversibility takes precedence, while in others, finality is more important and reversibility is undesirable. Other properties may be positive in an absolute sense, but their adoption forces the exclusion of other desirable properties. For instance, both untraceability and anti money laundering compliance could be considered desirable properties, but strengthening one will necessarily require weakening the other.
\ifnum \sver=0
Another such trade-off occurs in the context of quantum money: some \textit{private-key} quantum money schemes are unconditionally secure (e.g. Wiesner's scheme); whereas \textit{public-key} quantum money schemes require computational hardness assumptions~\cite{AC12}. As such, the choice between a private and a public key scheme implies a trade-off between the type of verifiability that the scheme supports and the level of security: public verifiability is preferred to private verifiability, but unconditional security is better than security based on computational assumptions. 
\fi

\ifnum \sver=0
\newcommand\tablewidth{0.95\columnwidth}
\else
\newcommand\tablewidth{0.90\linewidth}
\fi

\newcolumntype{g}{>{\columncolor{gray!25}}c}

\begin{table*}
    {
    \rowcolors{2}{gray!25}{}
    \setlength{\tabcolsep}{5pt}
    \resizebox{\tablewidth}{!}{\begin{tabular}{l|l|grgrgrgrgrgrg}
Primary Function & Property & \mcrot{1}{l}{60}{\twoelementtable{Commodity Money \cellcolor{gray!25}}{(Gold) \cellcolor{gray!25}}} &
\mcrot{1}{l}{60}{\twoelementtable{Physical coins}{(USA coins)}} & \mcrot{1}{l}{60}{\twoelementtable{Physical bills \cellcolor{gray!25}}{(USA bills) \cellcolor{gray!25}}} & \mcrot{1}{l}{60}{\twoelementtable{Central Bank Reserves}{(Bank reserves at the Federal Reserve)}} & \mcrot{1}{l}{60}{\twoelementtable{Bank deposits \cellcolor{gray!25}}{(USA bank savings account) \cellcolor{gray!25}}} & \mcrot{1}{l}{60}{\twoelementtable{CBDC}{(No mature realizations)}} & \mcrot{1}{l}{60}{\twoelementtable{In-game currency  \cellcolor{gray!25}}{(PokéCoin)  \cellcolor{gray!25}}} & \mcrot{1}{l}{60}{\twoelementtable{Cryptocurrency}{(Bitcoin)}} & \mcrot{1}{l}{60}{\twoelementtable{Cryptocurrency with DApps  \cellcolor{gray!25}}{(Ethereum)  \cellcolor{gray!25}}} & \mcrot{1}{l}{60}{\twoelementtable{Privacy oriented Cryptocurrency}{(Zcash)}} & \mcrot{1}{l}{60}{\twoelementtable{Stable Coin  \cellcolor{gray!25}}{(Tether ERC-20 USD) \cellcolor{gray!25}}} & \mcrot{1}{l}{60}{\twoelementtable{Private-Key Quantum Money }{(Wiesner's scheme [no realization])}} & \mcrot{1}{l}{60}{\twoelementtable{Public-Key Quantum Money  \cellcolor{gray!25}}{(Farhi et al. scheme [no realization])  \cellcolor{gray!25}}} \\
\midrule\cellcolor{white}\textbf{\hyperref[sec:medium_of_exchange]{Medium of Exchange\textsuperscript{$\dagger$}}} & \hyperref[par:acceptability]{Acceptability\textsuperscript{$\dagger$}} & \x & \chk & \chk & \x & \chk & \chk & \x & \minus & \minus & \minus & \minus & \chk & \chk \\
\cellcolor{white} & \hyperref[par:accessibility]{Accessibility} & \chk & \chk & \chk & \x & \minus & \chk & \minus & \chk & \chk & \chk & \minus & \chk & \chk \\
\cellcolor{white} & \hyperref[par:cognizability]{Cognizability\textsuperscript{$\dagger$}} & \chk & \chk & \chk & \chk & \chk & \chk & \chk & \chk & \chk & \chk & \chk & \x & \chk \\
\cellcolor{white} & \hyperref[par:digital]{Digital} & \x & \x & \x & \chk & \chk & \chk & \chk & \chk & \chk & \chk & \chk & \chk & \chk \\
\cellcolor{white} & \hyperref[par:divisibility_and_mergeability]{Divisibility\textsuperscript{$\dagger$} and mergeability} & \minus & \minus & \minus & \chk & \chk & \chk & \chk & \chk & \chk & \chk & \chk & \minus & \minus \\
\cellcolor{white} & \hyperref[par:ease-of-use]{Ease-of-use} & \x & \chk & \chk & \chk & \chk & \chk & \chk & \minus & \x & \minus & \minus & \chk & \chk \\
\cellcolor{white} & \hyperref[par:latency]{Latency} & \chk & \chk & \chk & \chk & \x & \chk & \chk & \minus & \chk & \minus & \chk & \chk & \chk \\
\cellcolor{white} & \hyperref[par:localVerifiability]{Local verifiability} & \chk & \chk & \chk & \x & \x & \x & \x & \x & \x & \x & \x & \x & \chk \\
\cellcolor{white} & \hyperref[par:computational_costliness]{Low computational tx cost} & \chk & \chk & \chk & \chk & \chk & \chk & \chk & \chk & \chk & \x & \chk & \chk & \chk \\
\cellcolor{white} & \hyperref[par:low_pecuniary_transaction_cost]{Low pecuniary tx cost} & \chk & \chk & \chk & \chk & \minus & \chk & \x & \volatile & \volatile & \chk & \volatile & \chk & \chk \\
\cellcolor{white} & \hyperref[par:P2P_transfer_mechanism]{P2P transfer mechanism} & \chk & \chk & \chk & \x & \x & \x & \x & \chk & \chk & \chk & \chk & \x & \chk \\
\cellcolor{white} & \hyperref[par:portability]{Portability\textsuperscript{$\dagger$}} & \minus & \minus & \minus & \chk & \chk & \chk & \chk & \chk & \chk & \chk & \chk & \chk & \chk \\
\cellcolor{white} & \hyperref[par:proof_of_payment]{Proof of payment} & \x & \x & \x & \chk & \chk & \chk & \x & \chk & \chk & \chk & \chk & \questionmark & \x \\
\cellcolor{white} & \hyperref[par:reputation]{Reputation} & \chk & \chk & \chk & \chk & \chk & \chk & \x & \minus & \x & \x & \x & \chk & \chk \\
\cellcolor{white} & \hyperref[par:reversibility]{Reversibility} & \x & \x & \x & \chk & \chk & \questionmark & \x & \x & \x & \x & \chk & \x & \x \\
\cellcolor{white} & \hyperref[par:smart-contracts]{Smart contracts} & \x & \x & \x & \x & \x & \chk & \x & \minus & \chk & \x & \chk & \x & \x \\
\cellcolor{white} & \hyperref[par:throughput]{Throughput} & \chk & \chk & \chk & \chk & \chk & \chk & \chk & \x & \x & \x & \x & \chk & \chk \\
\cellcolor{white} & \hyperref[par:transferability]{Transferability} & \chk & \chk & \chk & \chk & \chk & \chk & \x & \chk & \chk & \chk & \chk & \chk & \chk \\
\cellcolor{white} & \hyperref[par:Transparency]{Transparency} & \x & \x & \x & \chk & \chk & \chk & \x & \chk & \chk & \chk & \chk & \chk & \chk \\
\cellcolor{white} & \hyperref[par:untraceability]{Untraceability} & \chk & \chk & \chk & \x & \minus & \questionmark & \minus & \x & \x & \chk & \x & \minus & \chk \\
\hline\cellcolor{white}\textbf{\hyperref[sec:standard_of_deferred_payment]{Standard of Deferred Payment\textsuperscript{$\dagger$}}} & \hyperref[par:legal_tender]{Legal tender} & \x & \chk & \chk & \x & \x & \chk & \x & \x & \x & \x & \x & \chk & \chk \\
\hline\cellcolor{white}\textbf{\hyperref[sec:store_of_value]{Store of Value\textsuperscript{$\dagger$}}} & \hyperref[par:backup]{Backup} & \x & \x & \x & \chk & \chk & \chk & \chk & \chk & \chk & \chk & \chk & \x & \x \\
\cellcolor{white} & \hyperref[par:durability]{Durability\textsuperscript{$\dagger$}} & \chk & \chk & \chk & \chk & \chk & \chk & \chk & \chk & \chk & \chk & \chk & \minus & \minus \\
\cellcolor{white} & \hyperref[par:interest-bearing]{Interest-bearing} & \x & \x & \x & \chk & \chk & \chk & \x & \x & \chk & \x & \x & \x & \x \\
\cellcolor{white} & \hyperref[par:inside_or_outside]{Outside} & \chk & \chk & \chk & \chk & \x & \chk & \chk & \chk & \chk & \chk & \x & \chk & \chk \\
\cellcolor{white} & \hyperref[par:proof_of_reserves]{Proof of reserves} & \x & \x & \x & \chk & \chk & \chk & \x & \chk & \chk & \chk & \chk & \chk & \x \\
\cellcolor{white} & \hyperref[par:scarcity]{Scarcity\textsuperscript{$\dagger$}} & \chk & \chk & \chk & \chk & \chk & \chk & \chk & \chk & \chk & \chk & \chk & \chk & \chk \\
\cellcolor{white} & \hyperref[par:supply_measurability]{Supply measurability} & \x & \chk & \chk & \chk & \chk & \chk & \x & \chk & \chk & \minus & \chk & \chk & \chk \\
\cellcolor{white} & \hyperref[par:tax_evadability]{Tax evadability} & \chk & \chk & \chk & \x & \x & \x & \x & \chk & \chk & \chk & \questionmark & \x & \chk \\
\hline\cellcolor{white}\textbf{\hyperref[sec:unit_of_account]{Unit of Account\textsuperscript{$\dagger$}}} & \hyperref[par:cost_of_currency_exchange]{Cost of currency exchange} & \x & \x & \x & \chk & \chk & \chk & \x & \chk & \chk & \minus & \chk & \chk & \chk \\
\cellcolor{white} & \hyperref[par:fungibility]{Fungibility\textsuperscript{$\dagger$}} & \chk & \chk & \chk & \chk & \chk & \chk & \chk & \minus & \minus & \chk & \minus & \chk & \chk \\
\cellcolor{white} & \hyperref[par:stability]{Stability\textsuperscript{$\dagger$}} & \minus & \chk & \chk & \chk & \chk & \chk & \minus & \x & \x & \x & \chk & \chk & \chk \\
\hline\cellcolor{white}\textbf{\hyperref[sec:societal_or_regulatory]{Societal or Regulatory}} & \hyperref[par:AML-compliant]{AML Compliant} & \x & \x & \x & \chk & \chk & \chk & \chk & \x & \x & \x & \minus & \chk & \x \\
\cellcolor{white} & \hyperref[par:censorship_resistant]{Censorship resistant} & \chk & \chk & \chk & \x & \x & \x & \x & \chk & \chk & \chk & \x & \x & \chk \\
\cellcolor{white} & \hyperref[par:identity-based]{Identity-based} & \x & \x & \x & \chk & \chk & \chk & \minus & \x & \x & \x & \x & \chk & \x \\
\cellcolor{white} & \hyperref[par:public]{Public} & \x & \chk & \chk & \chk & \x & \chk & \x & \x & \x & \x & \x & \chk & \chk \\
\cellcolor{white} & \hyperref[par:resource-efficiency]{Resource efficiency} & \x & \chk & \chk & \chk & \chk & \chk & \chk & \x & \x & \x & \chk & \chk & \chk \\
\cellcolor{white} & \hyperref[par:unforgeability]{Unforgeability} & \chk & \chk & \minus & \chk & \chk & \chk & \chk & \chk & \chk & \chk & \chk & \chk & \chk
\\ \bottomrule
\end{tabular}}
\caption{Properties of money.}\label{tab:properties_of_money}
}
\small
The table categorizes instantiations of broad categories of money according to the extent to which they exhibit different properties. Each row contains a property of money, categorized by the primary function to which it corresponds. Each column refers to a broad category of money, along with a representative example, given in parentheses, and is used to determine which properties apply. A $\dagger$ indicates that a property or function appeared in the original Jevons-Menger framework. A \chk \  indicates that a form of money has a property, a \minus\ indicates that the property is present but weaker than in the best available implementations, an \x\ indicates that it is not present or not satisfactory, and A \questionmark \ indicates that we are  uncertain whether the property will hold. The \volatile\ symbol represents a volatile transaction cost. The precise definitions of less familiar forms of money are given in Appendix \ref{sec:appendix}. For the purpose of this table, we assume that quantum money is issued as a CBDC and, thus, has the properties of public money. It is also, of course, possible that it could be issued privately. In cases where there is lack of supporting information, we assume that the property is present if 1) it is trivially implementable with existing technology, 2) there are no binding legal or institutional constraints that prevent it from being implemented.
\end{table*}

\ifnum \sver=1
\subsection*{Medium of Exchange Function}
\else 
\subsection{Medium of Exchange Function}
\fi

\label{sec:medium_of_exchange}
Jevons~\cite{Jev76} describes a medium of exchange as something that is ``...esteemed by all persons... which any person will readily receive'' and a ``means of producing necessities of life at any time.'' As such, we may convert what we produce into a medium of exchange and then use that medium of exchange to purchase consumption goods. In this subsection, we examine properties of money that relate to its ability to function as a medium of exchange. 
\ifnum \sver=0
For a theoretical treatment of money's medium of exchange function, see Wallace~\cite{Wal80}, Kiyotaki and Wright~\cite{KW89}, Oh~\cite{Oh89}, Kiyotaki and Wright~\cite{KW93}, Williamson and Wright~\cite{WR94}, and Lagos~\cite{Lag10}. For experimental work on money as a medium of exchange, see Brown ~\cite{Bro96}, and Duffy and Ochs~\cite{DO99}.
\else
For a theoretical treatment of money's medium of exchange function, see ~\cite{Wal80,KW89,Oh89,KW93,WR94,Lag10}. For experimental work on money as a medium of exchange, see ~\cite{Bro96,DO99}.
\fi

\paragraph{Acceptability.} \phantomsection \label{par:acceptability} In order for money to function as a medium of exchange, it must be accepted as a form of payment. Menger~\cite{Men1892} observed that the liquidity of a good influenced its acceptability. This is why commodity money was a popular choice prior to the invention of fiat currencies: commodities were liquid and had intrinsic value, which made it less costly for merchants to accept them as a form of payment. In contrast, acceptability is a substantial limitation for cryptocurrencies and is an important consideration for CBDCs. For a discussion of the conditions that need to be satisfied for a new fiat currency to become ``acceptable,'' see Selgin~\cite{Sel94}. For a more theoretical treatment of acceptability, see ~\cite{KW92,SW04}.

\paragraph{Accessibility.} \phantomsection \label{par:accessibility}Bjerg \cite{Bje17} defined the concept of money ``accessibility'' as the answer to the question: Who can use this type of money? Bech and Garratt~\cite{BG17} use accessibility as one of the four criteria in their proposed taxonomy of money. Within this system, physical cash is considered to be ``universally accessible," since any person or entity may easily obtain and use it. To the contrary, central bank reserves are not universally accessible, since they are not available to the general public. We argue that private bank money also has limited accessibility -- relative to physical cash and cryptocurrencies -- since it is not easily accessible to some groups, such as minors and foreigners.

\paragraph{Cognizability.} \phantomsection \label{par:cognizability} Jevons \cite{Jev76} defines the cognizability of money as
\ifnum \sver=0
\begin{quote}
   the capability of a substance for being easily recognized and distinguished from all other substances. ... Precious stones, even if in other respects good as money, could not be so used, because only a skilled lapidary can surely distinguish between true and imitation gems.
\end{quote}
\else
``the capability of a substance for being easily recognized and distinguished from all other substances. ... Precious stones, even if in other respects good as money, could not be so used, because only a skilled lapidary can surely distinguish between true and imitation gems."
\fi

While the need for cognizability declined in the 20th century, it may once again become important in the emerging era of currency competition. Increased fragmentation in the monetary system, driven by a rise in the popularity of cryptocurrencies, other forms of private money, and competing CBDCs may make it difficult for users to identify counterfeit and fraudulent products. Indeed, theory suggests that reduced cognizability could result in increased counterfeiting and fraud~\cite{NW07, Wal10, Hu13}.

The quantum money literature provides a refinement to the concept of cognizability that requires that a unit of money that is verified once -- and, thus, appears to be valid -- must also pass further verification attempts. This rules out acts of sabotage, where the attacker harms others, but does not benefit monetarily\ifnum \sver=1 ~\cite{CS20,RS19proc,BS20}\fi. 

\ifnum \sver=0
To understand this refinement, consider the case where malicious Mallory has valid bills, but tampers with them in such a way that they pass verification on the first attempt, but fail on subsequent attempts. Mallory sends that money to honest Alice who accepts the bills after they pass the (first) verification attempt. However, when Alice attempts to spend the money, it will fail the (second) verification attempt. Therefore, Alice is harmed, even though Mallory does not gain anything directly from the attack. This standard of cognizability is difficult to achieve in the context of quantum money, where the outcome of verification is not necessarily deterministic~\cite{RS19proc,CS20,BS20,RZ20}.
\fi

\paragraph{Digital.} \phantomsection \label{par:digital} We define a form of money as ``digital'' if it can be exchanged remotely.\footnote{Bech and Garratt~\cite{BG17} propose a taxonomy of money where ``physicality'' constitutes one of the four core properties, and where physical is the opposite of digital.} Existing quantum money schemes fall into the ``digital'' category of money, since transactions are conducted through the exchange of information, rather than the exchange of physical tokens.

\paragraph{Divisibility and mergeability.} \phantomsection \label{par:divisibility_and_mergeability} 
Divisibility is typically interpreted as a relative measure. Lee and Wallace~\cite{LW06}, for instance, use the ratio $M/s$, where $M$ is the per capita money stock and $s$ is the size of the most common monetary unit. They find that this measure ranged from 25 to 130 in medieval Europe; whereas, it was closer to 40,000 for the United States in 2004.\footnote{Prior to the 19th century, money was not widely available in small denominations~\cite{Red00,SV02}. While may have been optimal~\cite{LW06}, it also likely that it impeded the functions of money.}

The concept of divisibility has been linked to money's medium of exchange function as early as Jevons~\cite{Jev76}, but gained renewed relevance in the era of digital currencies. Furthermore, its definition may also need to be revisited, since many forms of modern money, including bank deposits, can be divided into small denominations frictionlessly and without liquidity considerations. Thus, the most common denomination size may no longer be the relevant divisor.%
\ifnum \sver=0\footnote{Jevons \cite{Jev76} makes the following connection between money's divisibility and its capacity to serve as a medium of exchange: ``a minor inconvenience of barter arises from the impossibility of dividing many kinds of goods. A store of corn, a bag of gold dust, a carcase of meat, may be portioned out, and more or less may be given in exchange for what is wanted. But the tailor, as we are reminded in several treatises on political economy, may have a coat ready to exchange, but it much exceeds in value the bread which he wishes to get from the baker, or the meat from the butcher. He cannot cut the coat up without destroying the value of his handiwork. It is obvious that he needs some medium of exchange, into which he can temporarily convert the coat, so that he may give a part of its value for bread, and other parts for meat, fuel, and daily necessaries, retaining perhaps a portion for future use.''}
\fi

Demand deposits typically allow for divisibility down to the smallest denomination of coin or lower. Cryptocurrencies allow for an even higher degree of divisibility: the smallest denomination of Bitcoin is 1 satoshi, which is $10^{-8}$ bitcoins.
\ifnum \sver=1 
\footnote{An interesting use-case for such high divisibility is micro-payments (also called micro-transactions), which have been studied extensively in the context of electronic-cash systems~\cite{LO98, MR02}.
    One such example relates to the main challenge in file-sharing peer-to-peer networks, such as BitTorrent~\cite{Coh03}, which is to discourage free riding~\cite{JA05}. A simple way to encourage users to upload is to provide pecuniary incentives; however, this is challenging in a setting where there is no trust between the transacting parties. Fine-grained divisibility is useful because it permits payments for small chunks of data, eliminating the possibility of abusing the system.} The lightning network allows transactions as small as one millisatoshi, which is worth on the order of a millionth of a USD cent as of 2020.\footnote{See, e.g., \url{https://lightningwiki.net/index.php/Denominations}.}
\else
    An interesting use-case for such high divisibility is micro-payments (also called micro-transactions), which have been studied extensively in the context of electronic-cash systems.\footnote{See, e.g., ~\cite{LO98}, and \cite{MR02}, and the references therein.} One such example relates to the main challenge in file-sharing peer-to-peer networks, such as BitTorrent~\cite{Coh03}, which is to discourage free riding~\cite{JA05}. A simple way to encourage users to upload is to provide pecuniary incentives; however, this is challenging in a setting where there is no trust between the transacting parties. Fine-grained divisibility is useful because it permits payments for small chunks of data, eliminating the possibility of abusing the system.\footnote{For example, if a payment of 1 cent occurs after 1MB, a free rider might try to download 1MB of the file, and then ``run away" without paying. This free-rider might try to download the other parts of the file from other users, or other mechanisms. Naturally, in such systems, these types of strategies can be easily automated. If payment of a $10^{-6}$ cent is done after 1 byte, the user cannot download and ``run away" with more then 1 byte, which renders this attack useless.} There are existing solutions that claim to use an approach that is similar to the one mentioned here.\footnote{Tokens worth at least \$62 million USD that can be used in this market are currently in circulation as of January 2020. Evaluating the credibility of these services and tokens is outside the scope of this paper.}
\fi
In contrast, existing quantum money schemes would not allow for frictionless divisibility.

In addition to divisibility, forms of money differ in the extent to which units can be merged together, a property which we will refer to as mergeability. Gold, for instance, can both be divided into arbitrarily fine units and also merged by melting the pieces together.

Mergeability is always achievable by collecting multiple units of the same denomination. However, storing more units may require more resources. With respect to physical cash, mergeability can be measured as the minimum number of units required to sum to an arbitrary number, $x$. For instance, in an economy with an idealized form of cash, which comes in denominations of $10^k$ for $k\in \mathbb{N}$, the mergeability scaling would be at most $10 \cdot \lceil \log_{10}(x) \rceil$. 

Other forms of money, such as bank deposits and cryptocurrencies, achieve perfect scaling. That is, they do not require more resources to store more value.

\ifnum \sver=0
For example, in an economy that uses precious stones as commodity money, the amount of resources needed to store $N$ stones of a given size is higher than the amount needed for $N-1$. However, if multiple types of precious stones are used, this will naturally result in different ``denominations,'' allowing for the exchange of two low value stones for a higher value stone.
\fi

\paragraph{Ease of use.} \phantomsection \label{par:ease-of-use} The cognizability of money is closely related to its ease of use. While cognizability refers to the difficulty of determining whether a piece of money is valid, ease of use refers to the difficulty of conducting a transaction with a unit of money, part of which will involve determining whether it is valid. Survey evidence suggests that perceived ease of use may be an important factor in determining whether or not an individual is willing to use a new form of money, such as a cryptocurrency~\cite{GCL16}. 

\paragraph{Latency.} \phantomsection \label{par:latency} Latency is defined as the time it takes for a transaction to settle. There are several potential causes of increased latency. Physical constraints are one: the speed of light, for instance, could add an order of a second for every round of communication needed in a digital transaction. A Bitcoin transaction is confirmed only after it is mined in a block, which takes 10 minutes on average~\cite{Nak08}.
In other systems, such as credit cards, an inquiry into whether a transaction is fraudulent might incur a delay if the payer is asked to confirm the details of the relevant transaction. Note that latency is weakly coupled with the notion of reversibility.

\paragraph{Local verifiability.} \phantomsection \label{par:localVerifiability}
Local verifiability means that counterfeiting can be detected without the involvement of a trusted third party. It was introduced as one of the necessary properties of a public-key quantum money scheme by Aaronson \cite{Aar09}, but can also be used to evaluate the desirability of any form of money, including physical bills and coins, which can also be locally verified by checking markers of authenticity. In contrast, private bank money, private-key quantum money, and cryptocurrencies are \textit{not} locally verifiable, since they require communication either with an authenticator or a digital ledger. Jevons \cite{Jev76} argued that precious gems were not sufficiently cognizable, since counterfeits were difficult to detect without expertise, which can be viewed as an early evaluation of local verifiability. Finally, note that a scheme cannot have both local verifiability and backup; otherwise, it will be trivial to construct counterfeit bills that pass verification.

\paragraph{Low computational cost.} \phantomsection \label{par:computational_costliness} The main computational resources needed to participate in a transaction are the time complexity required to make and verify a transaction, network connectivity, liveliness, storage, memory, and power consumption. Informally, time complexity is the number of computational steps needed to perform or verify a transaction. And liveliness is the requirement that both participants to a transaction be online at least periodically.

Bank deposit transfers via contactless chip debit cards are an example of low computational costliness: all of the power needed is supplied over the air.%
\footnote{Chip cards are denoted Integrated Circuit Cards (ICCs) in the \href{https://www.emvco.com/emv-technologies/contactless}{EMV specification}.}
In contrast, the original Bitcoin client was computationally costly: users had to download the entire blockchain before they could send or receive bitcoins.\ifnum \sver=0\footnote{As of October 2021, it is more than 370GB, making it impractical for mobile phones. Source: \url{https://www.blockchain.com/charts/blocks-size}.}\fi This, of course, creates a burden for the users, both in terms of the storage requirements and network capacity. Already, in his original manuscript, Nakamoto suggested using a Simplified Verification Protocol (SPV), which would reduce both storage and network communication requirements. The security risks increase only slightly when one uses SPV wallets, rather than a full node. Indeed, SPV is  used in most of the recommended mobile Bitcoin clients today, and none of the mobile clients support full validation.\footnote{Source: \url{https://bitcoin.org/en/choose-your-wallet?step=5}.}

Private (shielded) transactions in ZCash have a high time complexity, storage and communication cost, since users have to download the full block-chain, store it, and perform an intensive computational task using that data~\cite{Pet18}.

\paragraph{Low pecuniary transaction cost.} \phantomsection \label{par:low_pecuniary_transaction_cost} 

Some payment instruments, such as bank transfers and credit cards, incur a pecuniary cost in the form of a fee imposed on the sender or receiver. To the contrary, transactions using physical cash do not.

Most cryptocurrencies require the sender to pay a transaction fee. This fee determines the priority with the order is handled. Especially during price surges, transaction fees tend to rise, as the demand for transactions rises. The fee structure differs from that of a credit card transaction, since it depends mostly on the total level of demand in the system, rather than on the amount of the transaction. This is similar to a check, which typically incurs a fixed fee per transaction that is not proportional to the amount.

Private-key quantum money schemes may, in principle, impose a validation fee, depending on the arrangement of the scheme. In contrast, public-key quantum money schemes do not require a third party to participate in the transaction, so no fee could be imposed, as is the case with physical cash.

\paragraph{P2P transfer mechanism.} \phantomsection \label{par:P2P_transfer_mechanism} Bech and Garratt~\cite{BG17} categorize forms of money by the mechanism used to transfer value. They define a peer-to-peer mechanism as one where ``...transactions occur directly between the payer and the payee without the need for a central intermediary.'' They also point out that ``On a computer network, the peer to-peer concept means that transactions can be processed without the need for a central server.''

\paragraph{Portability.} \phantomsection \label{par:portability} Portability is commonly referenced as a necessary property of money \cite{Jev76}. The need for portability is likely what lead to the creation of ``representative money,'' such as notes that could be converted into a commodity, to replace the use of the commodity itself as a form of money. While portability might appear to be trivially satisfied for all digital forms of money, it can be hard to achieve in certain cryptocurrency schemes. Until recently, ZCash shielded transactions required access to the entire blockchain, which created substantial storage requirements that could be prohibitive for mobile payments. 

\paragraph{Proof of payment.} \phantomsection \label{par:proof_of_payment} Suppose Alice pays \$1 to Mallory, the malicious merchant, to purchase a product. Mallory takes the \$1 bill for inspection, secretly replaces it with counterfeit money, and then passes the counterfeit bill to Alice, claiming that the money she paid with was invalid. This form of fraud cannot be done with other forms of payment. For example, with an (idealized) credit card service, there could be no such disagreement between Alice and Mallory, since the credit card company, which is assumed to be honest, serves as an intermediary. More generally, a ``proof of payment'' protocol can be used to prevent such disagreements. Bitcoin, for example, currently supports such a protocol~\cite{AH13}.\footnote{As far the authors are aware, the Bitcoin Lightning Network -- a second layer built on top of Bitcoin -- does not provide a proof of payment.}

On the other hand, public-key quantum money transactions leave no record and, similar to cash, do not offer an obvious means of achieving proof of payment. One possible workaround could be the following. Suppose Alice wants to send \$10 of quantum money to Mallory. Instead of sending it all at once, she could divide the payment into 1000 iterations. In each iteration, she would send $1$\textcent and expect a digital signature approving the payment in return. If Mallory fails to provide such a signature, Alice would abort. The worst case scenario is that Alice would not have a proof of payment for $1$\textcent. It is hard to imagine such a process being conducted with physical cash, but electronic forms of money could incorporate it at the protocol level, without most users even being aware of its existence. 

\paragraph{Reputation.} \phantomsection \label{par:reputation} One determinant of acceptability is the trust that users have in a form of money or in its issuer. As such, reputation or ``brand trust'' may provide valuable information about a form of money's capacity to function as a medium of exchange and has the advantage of being evaluable prior to issuance. CBDCs, for instance, may be evaluated in terms of a central bank's reputation for maintaining price stability. Similarly, privately-issued digital currencies, such as Facebook's Diem \cite{Lib21} may be evaluated in terms of the issuer's name recognition or reputation for technical prowess.

\paragraph{Reversibility.} \phantomsection \label{par:reversibility} In general, payment with physical forms of money, such as coins and bills, cannot be reversed unless both parties consent. This differs from digital forms of payment, such as private bank money, transferred via debit transaction, which allows for the reversal of transactions under certain circumstances. Allen et al. \cite{ACE+20} discuss a broader term, rectification, which also allows a user to correct information about themselves, and argue that a currency's rectifiability is typically increasing in the extent to which its ledger system is centralized.

With respect to quantum money, public-key schemes do not communicate with a trusted third party and do not leave a record. Thus, reversibility is not possible. Reversibility for private-key quantum money could be introduced, by allowing for an escrow period before settlement, hence, introducing a trade-off with latency. 

The reversibility of a CBDC will depend on the scheme that the central bank adopts. In general, schemes that are identity-based and centralized will afford a greater degree of reversibility.

\paragraph{Smart contracts.} \phantomsection \label{par:smart-contracts}

Buterin~\cite{But14} describes smart contracts as:
\ifnum \sver=1
``systems which automatically move digital assets according to arbitrary pre-specified rules. For example, one might have a treasury contract of the form `A can withdraw up to X currency units per day, B can withdraw up to Y per day, A and B together can withdraw anything, and A can shut off B's ability to withdraw.'"
\else
\begin{quote}
\textit{
... systems which automatically move digital assets according to arbitrary pre-specified rules. For example, one might have a treasury contract of the form ``A can withdraw up to X currency units per day, B can withdraw up to Y per day, A and B together can withdraw anything, and A can shut off B's ability to withdraw". (...) What Ethereum intends to provide is a blockchain with a built-in fully fledged Turing-complete programming language that can be used to create ``contracts" that can be used to encode arbitrary state transition functions, allowing users to create any of the systems described above, as well as many others that we have not yet imagined, simply by writing up the logic in a few lines of code.
}
\end{quote}
\fi

Bitcoin provides a scripting language which can be used to design simple smart contracts: notable examples include \textit{multi-sig(nature) transactions}, in which the consent of $m$-out-of-$n$ parties is needed to spend bitcoins; and \textit{atomic cross-chain swaps} \ifnum\sver=0(see page \pageref{par:atomic_swaps})\fi, which allow Alice and Bob to trade two cryptocurrencies without trusting each other~\cite{NBF+16}. 

More expressive platforms, such as Ethereum, allow for greater flexibility, which has enabled the development of decentralized finance (DeFi) and decentralized autonomous organizations (DAOs) \cite{But13, Ver18}; however, there are two main disadvantages to adopting an expressive platform: i) bugs, which are extremely hard to rule out in expressive platforms, could result in fraud or reputational damage; and ii) increased platform complexity, which arises in part from the difficulty of correctly calibrating fees to take into account the computational cost for miners or validators.

Allen et al. \cite{ACE+20} argue that CBDCs should not offer a scripting language to third-party developers for smart contracts, but should instead hardwire in a limited set of contracts to reduce the prevalence of bugs.
Quantum money does not solve the consensus problem or any variant of it, and existing constructions do not provide functionality for smart contracts. 

\paragraph{Throughput.} \phantomsection \label{par:throughput} In the context of payments, throughput measures the amount of transactions that can be processed in a system at a point in time. It is closely related to the concept of scalability in cryptocurrencies and other forms of digital payment \cite{ACE+20}. Physical cash faces no bottleneck that limits throughput. Credit card networks, in contrast, do face limitations, but have high rates of throughput. VISA and MasterCard, for instance, have claimed to be able to process 24,000 and 44,000 transactions per second, respectively~\cite{Her13}. In contrast, cryptocurrencies are typically low-throughput forms of payment. Bitcoin, for instance, processes at most 7 transactions per second and would need to change its protocol to substantially increase this rate~\cite{NBF+16}. Finally, public-key quantum money does not rely on any central bottleneck for verification and, thus, could achieve high throughput. 

\paragraph{Transferability.} \phantomsection \label{par:transferability} Transferability means that a form of money can be either physically or digitally transferred from one owner to another. Some historical forms of money, such as the large stones used on the Island of Yap \cite{Fur10, Fri91}, had transferable ownership, but could not be physically moved. Certain forms of in-game money, such as PokéCoin, may not be resalable and, thus, may either be spent or saved, but not transferred.

\paragraph{Transparency.} \phantomsection \label{par:Transparency} \onote{Go over again, especially discuss CBDC's. Update table.}Transparency can be either involuntary or optional. Since involuntary transparency is the opposite of untraceability, we will focus on the optional case. Zcash provides a clear example of this: users who want anonymity and privacy can have it; however, those who want transparency also have a mechanism for achieving it. Bitcoin also supports a similar notion called ``hierarchical deterministic wallets''~\cite{Wui13,NBF+16}. 

Optional transparency allows for the limited and voluntary exchange of information. This might include business partners who want to provide each other transparency with respect to their accounts, but do not want to provide such information to the public at large.

Physical cash does not produce records and cannot provide a mechanism for optional transparency. Bank accounts may offer a view-only permission to users that the account holder selects and, thus, can provide optional transparency. Most CBDC schemes could also provide a similar functionality.

\paragraph{Untraceability.} \phantomsection \label{par:untraceability}
Untraceability (or anonymity) makes it difficult to identify the users that are involved in a transaction \cite{CFN88}. Full untraceability is hard to achieve without \emph{privacy}, which entails hiding the existence and details of a transaction. This includes -- perhaps most importantly -- hiding the amounts involved \cite{APSX16}. 

Auer and Böhme \cite{AB20} and Allen et al. \cite{ACE+20} argue that there is a fundamental tradeoff in CBDC design between untraceability and anti money laundering compliance. Chaum et al. \cite{CGM21} propose a CBDC scheme that would allow for anonymity while still using the central bank as a trusted third party. Agur et al. \cite{AAD19} argue that CBDCs that focus on anonymity as a core property will tend to be substitutes for cash, rather than bank deposits. The decline of cash might increase the demand for digital forms of money that provide a strong form of untraceability.

Bitcoin has a low level of untraceability as a consequence of its ledger, which is open for everyone to inspect~\cite{RS13}. Some cryptocurrencies have improved upon Bitcoin by using various cryptographic techniques. These privacy-enhancing technologies for cryptocurrencies require various trade-offs. For example,  ZCash~\cite{BCG+14,HBHW16} requires a \textit{trusted-setup}, which reduces the level of unforgeability. Additionally, private or ``shielded'' transactions are four times bigger in size (2KB, instead of 0.5KB\footnote{Source: \url{https://z.cash/upgrade/}}). As of August 2021, the number of non-private transactions is an order of magnitude larger than the private transactions transactions.

In a private-key quantum money scheme, the central bank participates in every transaction and, thus, untraceability is not possible. Furthermore, quantum bills used in private-key schemes have unique (classical) serial numbers, so users do not have any privacy or anonymity with respect to the bank. In public-key quantum money schemes, the central bank itself is only involved during the minting and issuance of money, so it provides the same level of privacy and anonymity as physical cash.\footnote{A bill has a serial number, which can be used to perform tracking. Consequently, coins afford more privacy and anonymity.}

\ifnum \sver=1
\subsection*{Standard of Deferred Payment} \phantomsection
\label{sec:standard_of_deferred_payment}
\else 
\subsection{Standard of Deferred Payment} \phantomsection
\label{sec:standard_of_deferred_payment}
\fi

A standard of deferred payment is a broadly or legally accepted means of repaying debt. While it is often excluded from the list of functions of money, it was discussed in Jevons \cite{Jev76} and may provide a substantial advantage to public money in currency competitions.

\paragraph{Legal tender.} \phantomsection \label{par:legal_tender}

While private bank money is a \textit{de facto} acceptable means of discharging debt and paying taxes, central bank issued currencies are typically the only form of money that is \textit{de jure} acceptable and, thus, ``legal tender.'' While having legal tender status is likely to improve the acceptability of a form of money, it does not necessarily imply that it must be legally accepted as payment for goods and services \cite{Fed21}.

\ifnum \sver=1
\subsection*{Store of Value Function} \phantomsection \label{sec:store_of_value}
\else 
\subsection{Store of Value Function} \phantomsection \label{sec:store_of_value}
\fi

Jevons \cite{Jev76} and Menger \cite{Men1892} both argued that one function of money was to act as a store of value. That is, goods and services can be converted into money, stored for a period of time, and then converted into other goods and services for the purpose of consumption. This allows producers of perishable goods to sell them immediately and store their value in a medium that does not rapidly depreciate. For a discussion of the store of value property in the economics literature, see \cite{Tob65}, \cite{Dav72}, and \cite{Wei87}.

\paragraph{Anti-theft prevention.} \phantomsection \label{par:anti_theft_prevention}

Forms of money differ in their capacity to prevent theft. Since theft prevention will depend on the use of best practices and, thus, will vary across users, we do not attempt to rank it across forms of currency. In general, applying the best practices when safeguarding digital forms of money will allow for a high standard of security while having a minimal impact on the money's capacity to carry out its functions. Compare, for instance, the use of one-time passwords for digital money to lockbox banking for commodity money. It is clear that the latter more substantially inhibits money's medium of exchange function. With respect to cryptocurrencies and CBDCs, key management is a particularly important dimension of anti-theft security. Secure hardware, such as hardware wallets, can be used as an effective anti-theft mechanism, but should not be the basis for achieving \hyperref[par:unforgeability]{unforgeability} in a currency \cite{ACE+20}.

\paragraph{Backup.} \phantomsection \label{par:backup} The ability to back up a form of money provides protection against computer failure and loss. Cash and coins cannot be backed up, since they are physical tokens that are not traceable to an individual. Cryptocurrencies, such as Bitcoin, can be backed up by saving a private key. There are, of course, many practical aspects of a good backup system. 

Bitcoin, as well as many other cryptocurrencies, support a standardized mnemonic based system~\cite{PRV+13}, and provide the following motivation for it:
\begin{quote}
\textit{A mnemonic code or sentence is superior for human interaction compared to the handling of raw binary or hexadecimal representations of a wallet seed. The sentence could be written on paper or spoken over the telephone.}
\end{quote}
In addition, Bitcoin supports a passphrase that would be needed to access the backup~\cite{CV12}. This adds another layer of security to withstand, for example, an ``evil maid attack'' from an adversary who gains physical access to the backup. 

In contrast, public-key quantum money schemes cannot be backed up, since the quantum states underlying the money cannot be copied and there is no public record of transactions. In principle, a central bank could put a mechanism in place to provide backup for private-key quantum money, such as the scheme introduced in Coladangelo~\cite{Col19}.

\paragraph{Durability.}\phantomsection \label{par:durability} Prior to the development of metallic coins, the durability of a form of money was an important consideration. Jevons \cite{Jev76} identifies corn in ancient Greece, olive oil in the Mediterranean, and jewelry in pre-colonial North America as goods that were sufficiently durable to fulfill the functions of money. All digital forms of money that offer a form of backup satisfy a higher standard of durability than is achievable with any physical currency. Those without backup, including quantum money, are as durable as the device on which they are stored. See Taub \cite{Tau85} for a theoretical analysis of durability in the context of commodity money.

\paragraph{Interest-bearing.} \phantomsection \label{par:interest-bearing} 
Physical cash is not associated with an account or a record of ownership and, thus, cannot be interest-bearing; however, competing forms of money, such as bank deposits, cryptocurrencies, and CBDCs do not have an equivalent limitation and could, in principle, bear interest. As Brunnermeier and Niepelt~\cite{BN19} discuss, constructing a CBDC with an interest rate gives the central bank another tool for conducting monetary policy. Furthermore, negative interest rates on CBDCs could be used to extend the effective lower bound (ELB) if physical cash eventually disappears due to lack of demand \cite{BG17,Ben19}, which could be useful during slow recoveries \cite{AK19}.

An interest-bearing CBDC could also have negative implications for financial stability by disintermediating the financial sector \cite{KK19}; however, some argue that a well-designed CBDC could avoid this \cite{And20}.\footnote{Thus far, central banks have been hesitant to propose CBDC instantiations that include an interest rate, which may indicate that substantial concerns about disintermediation remain.} Garratt and Zhu \cite{GZ21} argue that an interest-bearing CBDC would put a lower bound on deposit rates, forcing larger banks to increase rates to compete when they could otherwise rely on network effects to lock-in customers. George et al. \cite{GXA21} argue that having the option to adjust the rate on a CBDC would allow the central bank to achieve monetary autonomy and exchange rate stability.

\paragraph{Outside (or inside).} \phantomsection \label{par:inside_or_outside} Forms of money are said to be either ``inside'' or ``outside.''\footnote{Alternatively, inside money is sometimes called ``private''; whereas, outside money is called ``public.''} Inside money, such as private bank money, is an asset for the holder and a liability for the issuer. Outside money, such as central bank-issued fiat currency, is an asset for the holder, but is not the liability of any private entity. The distinction between inside and outside money has seen increasing attention in the literature recently, as economists have attempted to describe the properties of new forms of money \cite{BG17, ZH19, BN19, BJL19}. Most existing quantum money schemes, including both private and public schemes, involve a trusted third party to perform issuance and -- for private-key schemes -- to perform verification. That entity is typically assumed to be a central bank, but, in principle, could be a private company or organization. As such, quantum money could be produced as either inside or outside money. Note that we use a checkmark in Table \ref{tab:properties_of_money} to indicate that a form of money is outside money, but do not take an stance on the desirability of the property.

\paragraph{Proof of reserves.} \phantomsection \label{par:proof_of_reserves}
This property allows participants (typically exchanges) to attest that they have some reserve that surpasses their liabilities. In cryptocurrencies, this is done by digitally signing a message using all the private keys in their control~\cite[Section 4.4]{NBF+16}. A similar functionality could be achieved with bank deposits where a customer could prove her reserve to others by showing her digitally signed bank statement. In the above two examples, note that cheating is still possible by colluding with others who are in control of the money (e.g., by borrowing temporarily).

\paragraph{Scarcity.}\phantomsection \label{par:scarcity}
Scarcity is defined in the context of social wealth by Walras~\cite{Wal26} as: ``All things, material or immaterial ... that are useful to us and ... only available to us in limited quantity.'' The scarcity of commodity money primarily refers to its natural abundance and the cost of extracting additional units. The scarcity of fiat currencies, including CBDCs and CBDC-based quantum money, is determined by the central bank's supply rule and the difficulty of counterfeiting. Cryptocurencies, such as Bitcoin, are sufficiently scarce to satisfy this definition.

\paragraph{Supply measurability.} \phantomsection \label{par:supply_measurability} Some forms of money, such as cryptocurrencies, can provide accurate measurements of the amount of money in circulation. Central banks also periodically provide measurements of the amount of physical bills and coins minted and put into circulation. In addition to this, some cryptocurrencies are also able to provide information about the projected future path of supply.\footnote{The time-inconsistency problem makes such commitments difficult for central banks, since it would sometimes require implementing an undesirable policy at a future date \cite{MJH17}.} For example, with Bitcoin, an adversary with the majority of the hashing-power could steal other's people money in certain cases; however, even such an adversary cannot issue more than 21 million bitcoins.

Supply measurability can be hard to achieve, especially in privacy oriented cryptocurrencies, due to ``hidden inflation'' -- an unrecorded increase in the money supply -- which could occur due to invalid computational assumptions or bugs in the code.\footnote{Note that ``inflation'' here refers to the amount of money in circulation, rather than the growth rate of the price level.}
This is not only a theoretical risk: such a flaw occurred in the implementation of ZCash (see Supplementary Material ~\ref{sec:appendix}).\footnote{See \url{https://electriccoin.co/blog/zcash-counterfeiting-vulnerability-successfully-remediated}.} Interestingly, there is no definitive way to know whether that bug was exploited, and therefore, what is the total supply of ZCash. In Ethereum, the amount of ether in circulation is known, but there are only few guarantees regarding future amounts.\footnote{See \href{https://github.com/ethereum/EIPs/blob/master/EIPS/eip-1559.md\#eth-burn-precludes-fixed-supply}{EIP-1559}.}

\paragraph{Tax evadability.} \phantomsection \label{par:tax_evadability} From a user perspective, forms of money that do not facilitate the assessment and collection of taxes may be considered more desirable. In the services industry, for example, employees may prefer to receive tips in the form of physical cash to avoid creating a paper or electronic trail that could be used to impose taxes. Additionally, foreign investors operating in a country with a history of financial repression may want to ensure that their funds are not subject to surprise taxes or confiscation. A form of money's tax evadability is positively related to its untraceability and censorship resistance, and inversely related to its level of AML compliance. This tension is an instructive example of the distinction between monetary properties that facilitate functions that provide private value and those that achieve a societal or regulatory function.

\ifnum \sver=1
\subsection*{Unit of Account Function} \phantomsection \label{sec:unit_of_account}
\else 
\subsection{Unit of Account Function} \phantomsection \label{sec:unit_of_account}
\fi

Jevons \cite{Jev76} argues that the unit of account or ``common measure of value'' function of money typically arises as a consequence of its use as a medium of exchange: ``Being accustomed to exchange things frequently for sums of money, people learn the value of other articles in terms of money, so that all exchanges will most readily be calculated and adjusted by comparison of the money values of the things exchanged.'' Brunnermeier et al. \cite{BJL19} argue that the functions of money may become unbundled in the digital era, such that one form of money may serve as the unit of account while rarely being used as a medium of exchange. In this subsection, we will discuss the properties of money that relate to its ability to function as a unit of account.\footnote{For a modern treatment of the unit of account function of money, see \cite{Mus77, Whi84, DS17}.}

\ifnum \sver=0
In digital settings, prices could be presented according to the unit preferred by the user, and exchanging can be done automatically on the merchant's side. In digital forms of money, switching costs may be sufficiently low that they are not an important consideration in determining which currency to hold. In this case, the need for a currency to be a unit of account is diminished \cite{BJL19}.
\fi

\paragraph{Cost of currency exchange.} \phantomsection \label{par:cost_of_currency_exchange} Every form of money has a cost associated with its exchange into other currencies. This includes the pecuniary costs incurred by the currency exchange, and the pecuniary and non-pecuniary costs incurred by users. Such costs are higher for some forms of money than others, since the difficulty of exchanging currencies is not uniform. In general, exchanges that are digital, involve highly liquid currencies, and entail minimal risk of fraud will tend to have lower costs.

Dyhrberg et al. \cite{DFS18} evaluate the transaction costs and liquidity of Bitcoin. They find that the quoted spreads across the Gdax, Gemini, and Kraken marketplaces average 5.60 to 22.51 basis points (bps). This is considerably lower than spreads in equity markets, but higher than spreads on commonly traded fiat currencies.

A low cost of currency exchange will tend to enhance a currency's capacity to act as a unit of account, since the unit of account will need to be exchanged frequently.

\paragraph{Fungibility.}\phantomsection \label{par:fungibility} The notion of fungibility is defined in McCloskey~\cite{McC16} as ``... a Latin legal term meaning `such that any unit is substitutable for another' ... A debt can be discharged with any money, not merely moneys from a particular account.'' Private bank money, central bank reserves, physical cash, and cryptocurrencies all appear to be fungible; however, as Poelstra et al. \cite{PBF+18} have argued, this is not as it seems. For instance, different units of Bitcoin contain different exchange histories that are traceable and may be undesirable. Bitcoin exchanges have even blocked the transfer of Bitcoins that originate with theft. In contrast, units of Zcash can be made indistinguishable and, thus, may be considered to be fungible. The greater capacity for non-fungibility in digital currencies could be seen as a positive property that can facilitate the distribution of helicopter drops, government benefits, and loans \cite{ACE+20}.

\paragraph{Stability.} \phantomsection \label{par:stability}
Black et al. \cite{BHM09stability} defines price stability as ``...maintaining the rate of increase or decrease in an aggregate price index, usually the consumer price index, within tolerable limits.''

Fiat currencies maintained by independent, inflation-targeting central banks have largely achieved price stability since the 1990s \cite{Sve10}. 

In contrast, cryptocurrencies, such as Bitcoin, have notoriously suffered from a lack of stability as a consequence of their supply rules~\cite{G719}. This gave rise to demand for cryptocurrencies with low price volatility, referred to as ``stablecoins'' \cite{HHS18}. Stablecoins rely on one of two mechanism to achieve parity with a target currency: 1) an algorithmic supply rule or 2) a guarantee of convertibility into some asset~\cite{CGM21,MIO+19}. Thus far, asset-backed stablecoins, such as Tether, have demonstrated a greater capacity for achieving price stability \cite{Coh19,CGM21}.\footnote{Adrian and Mancini-Griffoli \cite{AM19} discuss how a public-private partnership could improve asset-backed stablecoins further by using central bank reserves as the underlying asset.}

Quantum money does not rely on the use of a distributed ledger and could be issued as a retail CBDC. As such, it could achieve stability properties that are similar to existing fiat currencies.

\ifnum \sver=1
\subsection*{Societal or Regulatory Functions} \phantomsection
\label{sec:societal_or_regulatory}
\else 
\subsection{Societal or Regulatory Functions} \phantomsection
\label{sec:societal_or_regulatory}
\fi

In addition to the original functions introduced in Jevons \cite{Jev76} and Menger \cite{Men1892}, forms of money in the digital era have increasingly begun to embody explicit societal and regulatory objectives. Such functions are not necessarily intended to improve user experience and may even make it worse. We consider such functions in this section.

\paragraph{Anti-money laundering (AML) compliant.}\phantomsection \label{par:AML-compliant} Levi and Reuter \cite{LR06} define money laundering as ``techniques for hiding proceeds of crime [which] include transporting cash out of the country, purchasing businesses through which funds can be channeled, buying easily transportable valuables, transfer pricing, and using underground banks.'' Anti-money laundering is a ``routinized set of measures to affect criminal revenues passing through the financial system.''

According to Allen et al. \cite{ACE+20} anti-money laundering (AML) measures are typically based on three types of laws. The first makes money laundering illegal, whether or not the act it conceals is illegal. The second creates reporting requirements for financial institutions, such as Know Your Customer (KYC) rules, which are intended to detect and hinder money laundering. And the third makes it illegal to attempt to circumvent such reporting requirements.

Forms of money vary in their capacity to achieve AML compliance. We define an AML compliant currency as having two properties: 1) the capability to detect and record illicit financial transfers; 2) the technical or administrative capacity to perform detection and reporting. Cash and bank deposits are examples of forms of money with weak and strong AML compliance properties, respectively. Cryptocurrencies, such as Bitcoin, exceed the capacity of even bank deposits along criterion (1), but lack an authority or mechanism for criterion (2).

\ifnum \sver=0
Money laundering sometimes falls under the umbrella of ``financial crimes.'' We concentrate on money laundering specifically because it is the most studied financial crime in the literature and the techniques used to prevent money laundering are similar to those used to prevent financial crime more generally. A form of money's capacity to collect taxes and enforce liens \cite{ACE+20} could be considered closely related properties.
\fi

\paragraph{Censorship resistant.} \phantomsection \label{par:censorship_resistant}
Some governments censor certain forms of online communication. In censorship resistant systems, such acts are challenging by design~\cite{KES+16}. In our context, censorship could take the form of confiscating money or banning transactions (for example, by dissidents). Forms of money in which trusted third parties are involved, such as bank deposits and certain stable coins\footnote{E.g., Tether has a black-listing mechanism, see Lines 268--305 in their~\href{https://etherscan.io/address/0xdac17f958d2ee523a2206206994597c13d831ec7\#code}{code}.} are easier to censor compared to those that do not involve third parties, such as cryptocurrencies~\cite{Wea18}, cash, and public-key quantum money. Allen et al. \cite{ACE+20} argue that censorship resistance is typically an increasing function of the extent to which a form of digital money is decentralized.

\paragraph{Identity-based.}\phantomsection \label{par:identity-based} Identity-based forms of money, such as bank deposits, require participants to use their true identities; whereas other forms of money, such as cash, do not. For digital money, the requirement to reveal one's identity often arises as a result of \hyperref[par:AML-compliant]{AML compliance}. Identity-based systems also have the advantage of allowing for the cultivation of an individual's reputation. Credit rating is an example of such a reputational mechanism. A variety of different protocols can be used for identity verification, including in-person verification, online verification, proxies, biometric markers, and social trust networks \cite{ACE+20}.

A closely related and more common division in the economics literature is the distinction between ``account-based'' and ``token-based'' forms of money. Demand deposits, for instance, are a type of account-based money. According to Kahn and Roberds \cite{KR09}, who provide an overview of the economics of payment systems, an account-based system must employ two technologies. The first records all actions taken by an account owner and the second verifies accounts. To the contrary, token-based money -- sometimes referred to as ``value-based'' or ``store-of-value'' money -- relies exclusively on a technology that can be used to verify the validity of a given token, such as a commodity or a unit of fiat currency \cite{KR09}. We do not adopt this definition because it does not distinguish between most modern forms of digital money, such as cryptocurrencies and proposed instantiations of CBDCs.

\paragraph{Public.} \phantomsection \label{par:public} Public money is any form of money that is issued by a government entity. This includes central bank-issued bills and coins, government bonds, and CBDCs. Some have argued that a transition from private money to public money (e.g. private bank money to a CBDC) would result in a credit crunch. Brunnermeier and Niepelt \cite{BN19} show that this is not necessarily true and identify the conditions under which the equilibrium allocations would be identical after a swap from private to public money.

\paragraph{Resource efficiency.} \phantomsection \label{par:resource-efficiency} 
Forms of money differ with respect to the costliness of issuance and maintenance. Public forms of money, such as central bank issued cash and coins, are arguably less efficient than private bank money, but considerably more efficient than commodity money or cryptocurrencies.\footnote{For commodity money and other forms of currency that can be legally mined or minted, we expect the marginal cost of production to be close to the price of the commodity or money.} According to the U.S. Federal Reserve System, for instance, minting a \$100 note costs just 14 cents.\footnote{For a \$100 bill, this amounts to 0.14 cents per dollar. Producing a \$1 note costs 6.2 cents.} Furthermore, the entire cost of currency operations at the Board of Governors was less than 1.1 billion USD in 2021.\footnote{See \url{https://www.federalreserve.gov/faqs/currency_12771.htm} for an overview of minting costs for different denominations of U.S. currency.} In contrast, bitcoin mining -- the process which prevents double spending and mints new bitcoins -- is estimated to account for 0.46\% of worldwide electricity consumption as of October 2021.\footnote{See estimates from the Cambridge Centre for Alternative Finance \url{https://cbeci.org/}} Alternatives to Bitcoin's consensus mechanisms that do not require mining have been both proposed and implemented~\cite{XZL+20}.

\paragraph{Unforgeability.}\phantomsection \label{par:unforgeability} Forms of money differ with respect to the security scheme employed and, consequently, the level of protection afforded against counterfeiting. Physical cash employs special threads and inks that are difficult to replicate. For the U.S., for instance, Quercioli and Smith \cite{QS15} find that counterfeits account for roughly 1 out of every 10,000 bills. In contrast, private bank money relies on cryptographic schemes that make computational assumptions about potential attackers, which may be rendered ineffective by advances in algorithms or hardware. Certain existing forms of encryption, such as RSA, may eventually become vulnerable to attacks from quantum computers, which can perform prime factorization almost exponentially faster than classical computers, using Shor's algorithm \cite{Sho94}. Bitcoin is also known to be susceptible to quantum attacks \cite{ABL+18}.\footnote{In the context of digital currencies, Allen et al. \cite{ACE+20} argue that the structure of the digital ledger determines the protection a form of digital money can provide against counterfeiting attempts. Common architectures include full decentralization, role separation, trust dispersal, and threshold trust. See \cite{ACE+20} for definitions of the terms.}

Private-key quantum money, including Wiesner's original scheme \cite{Wie83}, achieves ``information-theoretic security,'' which means that an attacker with unbounded computational resources would still be unable to counterfeit a unit. Here, we assume the adversary receives $k$ valid money states from the bank, applies an arbitrary (perhaps inefficient) quantum computation with these states, and submits $m=poly(n)$ alleged money states to the bank, where $n$ is the number of qubits of the money state. We say that the scheme is secure if the probability of the adversary to pass $k+1$ or more verifications is negligible in $n$.  Perhaps surprisingly, full security proofs for Wiesner's money were given only 3 decades later~\cite{PYJ+12,MVW12}.

Finally, similar to debit card transactions, public-key quantum money schemes must rely on computational assumptions~\cite{AC12}, and are not information-theoretically secure. For example, the construction in Farhi et al. \cite{FGH+12} relies on the assumption that a certain computational problem in knot-theory is intractable to quantum computers. The reason is essentially as follows: a computationally unbounded adversary can enumerate over all quantum states (up to some precision $\epsilon$) and check whether it passes verification. Since the verification procedure is public, this does not require any cooperation from the bank. Notice that the same approach would not work for private money, since the bank would accept only polynomially-many states from the adversary for verifications; whereas brute-force attacks, such as this one, would require exponentially many attempts. This is essentially the same reason why guessing a short random password takes an exponentially long time.

\ifnum \sver=0
\section{Properties of Currency Pairs and Groups}

In some cases, properties of money extend to a pair of currencies or a group of currencies. We consider five such properties in this section. For the sake of simplicity, we do not categorize pairwise and group properties according to function, and do not attempt to determine whether they apply to each combination of currencies in Table \ref{tab:properties_of_money}.

\paragraph{Atomic swaps.} \phantomsection \label{par:atomic_swaps} The vulnerability of cryptocurrency exchanges to hacking \cite{Tho20, Sel20} has given rise to demand for an intermediary-free form of cryptocurrency exchange. A technique called an \textit{atomic swap} enables such exchanges between cryptocurrencies through the use of smart contracts.~\cite[Chapter 10.5]{NBF+16}. Such technology could also potentially be used in CBDCs to allow for peer-to-peer foreign currency exchange.

\paragraph{Interoperability.} \phantomsection \label{par:interoperability} In the context of cryptocurrencies, interoperability refers to the existence of protocols that allow two independent digital ledger systems to interact through the use of smart contracts \cite{LXS+19, GRS16, ACE+20}. Allen et al. \cite{ACE+20} propose a notion of interoperability that would allow for a two-layer CBDC. The central bank would manage a layer that corresponds to reserves and has only basic functionality. Commercial banks would then manage a retail layer that contains more customer-centric functionality, but is ultimately backed by holdings in the reserve layer.

\paragraph{Cross-border payments.} \phantomsection \label{par:cross_border_payments} Allen et al. \cite{ACE+20} argue that private digital currencies, such as cryptocurrencies, may improve the efficiency of cross-border payments, since they can potentially improve tracking and can eliminate the need for multiple financial intermediaries to be involved in a transfer. The BIS has also argued that improvements in cross-border payments could be facilitated by making CBDCs interoperable \cite{BIS21}.

\paragraph{Uniformity.} Uniformity between multiple currencies can be achieved by guaranteeing convertibility at a fixed rate between one currency and another. This allows one currency to take on another currency's store of value and unit of account properties. Brunnermeier et al. \cite{BJL19} point out private bank money (e.g. demand deposits) as an example of a currency that achieves this property. They also argue that the issuance of CBDCs could extend public money to substantially larger group, ensuring the uniformity of money in the era of digitization. Since quantum money could also be issued by central banks, possibly as a form of CBDC, it could also achieve uniformity with physical cash and private bank money. Stablecoins which are pegged to a single currency, such as the US dollar, may also achieve uniformity.
\fi

\ifnum \sver=1
\section*{Discussion}
\else 
\section{Discussion}
\fi

After an extended period of dominance in the 20th century, national forms of public money have fallen out of favor as a medium of exchange, losing market share to private bank money, even as they retain their status as the preferred unit of account. In the emerging era of intense digital currency competition, central banks have the opportunity to regain control over the medium of exchange through CBDC issuance, but face the threat of losing control over the unit of account to a multi-currency stablecoin, a competing central bank, a digital currency area \cite{BJL19}, or a cryptocurrency. Such an event would have substantial implications for monetary policy, financial stability, and regulation. As such, the conservative inclinations of central banks, which normally play a stabilizing role, could instead lead to loss of relevance for public money.

The emergence of new forms of public and private money raises questions about what properties of money are most beneficial in the modern era. Central banks appear to have concluded that new forms of public money need to be digital, but beyond that, there is less agreement on what other properties are desirable. Furthermore, the use of a digital medium opens up the possibility of embedding new supervisory and regulatory functions into money, which may be desirable from a societal perspective, but not from the perspective of an individual user. Our intention in this paper was to provide an overview of this emerging landscape that updates the functions-and-properties framework of money, and that could be useful for both researchers and currency designers.
 
\ifnum \sver=1
\showmatmethods{}

\acknow{O.S. is supported by the Israeli Science Foundation (ISF) grant No. 682/18 and 2137/19, and by the Cyber Security Research Center at Ben-Gurion University.}

\showacknow{}
\else 
\paragraph{Acknowledgments}
O.S. is supported by the Israeli Science Foundation (ISF) grant No. 682/18 and 2137/19, and by the Cyber Security Research Center at Ben-Gurion University.
\fi

\ifnum \sver=1 
    \bibliography{References}
    \pagebreak
\fi

\appendix

\section{Appendix} \label{sec:appendix}

Below, we provide additional detail about some of the potentially less familiar currencies listed in Table \ref{tab:properties_of_money}.

\paragraph{CBDC (No mature instantiation).} \phantomsection \label{def:cbdc} 
Since CBDCs lack a mature instantiation, we considered a generic case where the properties were determined by hard technical constraints, and common legal and institutional restrictions on central banks.

\paragraph{In-Game Currency (PokéCoin).} \phantomsection \label{def:pokecoin} In-game currency is one of the main drivers of game mechanics. Spending on mobile games alone has been estimated to have reached \$79 billion globally in 2020.\footnote{See \url{https://sensortower.com/blog/app-revenue-and-downloads-2020}.} PokéCoin is the virtual in-game currency used in the Pokémon GO game.
\paragraph{Cryptocurrency (Bitcoin).} \phantomsection \label{def:bitcoin} Bitcoin is an electronic payment system~\cite{Nak08,NBF+16}. Its censorship resistance and global accessibility are achieved mainly by its p2p architecture. Unlike previous e-payment systems, the Bitcoin network was the first to issue its own form of outside money via a process called mining. Mining provides the distribution mechanism of newly minted bitcoins, but even more importantly, plays a crucial role in securing the network against double-spending attacks. 

\paragraph{Cryptocurrency with DApps (Ethereum).} \phantomsection \label{def:ethereum}
Decentralized applications (DApps) are software that can be executed on the blockchain. DApps have enabled the development of decentralized financial services, which allow for lending and borrowing without an intermediary (other than the blockchain). Ethereum is a cryptocurrency oriented towards \hyperref[par:smart-contracts]{smart contracts}.

\paragraph{Privacy Oriented Cryptocurrency (Zcash).} \phantomsection \label{def:zcash}
Zcash is a cryptocurrency that allows users to enhance the privacy of transactions~\cite{HBHW16}. Each transaction is either ``transparent'' or ``shielded.'' The transparent transactions provide a level of privacy that is similar to Bitcoin; whereas the shielded transactions use a cryptographic protocol that involves zero-knowledge proofs to provide enhanced anonymity and privacy for transactors. Gross et al. \cite{GSB+21} propose the use of a Zcash-like shielding mechanism in a CBDC.

\paragraph{Stablecoin (Tether ERC-20 USD).} \phantomsection \label{def:tether}

A stablecoin is a type of cryptocurrency that attempts to achieve reduced price volatility. Tether ERC-20 USD is a stablecoin that is pegged to the value of the U.S. dollar (USD). Tether Limited, which issues the cryptocurrency, claims that its tokens are fully backed by USD reserves. ERC-20 is a protocol that is used for the Ethereum network. A unit of Tether ERC-20 USD is a token, which can be exchanged over the Ethereum blockchain.

\paragraph{Private-Key Quantum Money (Wiesner's scheme [no realization]).} \phantomsection \label{def:private_quantum_money}
The first quantum money scheme was introduced by Wiesner \cite{Wie83}. 
In a private-key scheme, only the bank branches or the central bank can verify the money, using the bank's secret key (vis-à-vis \hyperref[def:public-quantum-money]{Public-Key Quantum Money}).

Wiesner's scheme uses the No-Cloning Theorem \cite{WZ82} to construct physically unforgeable money, something which is not possible without exploiting quantum phenomena. The scheme was partially implemented in a laboratory setting by Bozzio et al. \cite{BOV+17}, but faces substantial technical barriers to a full implementation. See Hull et al. \cite{HSDW20} for a complete description of Wiesner's scheme.

\paragraph{Public-Key Quantum Money (Farhi et al. scheme [no realization]).} \phantomsection \label{def:public-quantum-money}

Public-key quantum money, a term introduced in \cite{Aar09}, refers to quantum money schemes with a publicly available key (algorithm) that can be used to perform counterfeiting detection. This would allow for local verification of money without the involvement of a trusted third party, something which is not possible with digital forms of classical (non-quantum) money. The scheme by Farhi et al. \cite{FGH+12} is based on knot theory. See Hull et al. \cite{HSDW20} for a summary of the scheme.

\section{Figures}\label{sec:figures}

\ifnum \sver=0
    \bibliographystyle{pnas-new-mod}
    {\footnotesize \bibliography{References} }
\fi 
\end{document}